{}
{}

\documentclass[11pt,a4paper]{article}

\newif\ifpublic\publictrue

\usepackage{mathtools,mathrsfs,amsbsy,amssymb,latexsym,amsfonts,amscd,amsmath}
\usepackage[nosort]{cite}
\usepackage{graphicx}
\usepackage[usenames,dvipsnames]{color}
\usepackage[normalem]{ulem}
\usepackage{pdfpages}
\definecolor{linkcolor}{rgb}{0,0,0.6}
\usepackage[colorlinks=true,pdfstartview=FitV,linkcolor=linkcolor,
citecolor=linkcolor,urlcolor=linkcolor,hyperindex=true,hyperfigures=false]{hyperref}
\ifpublic\else\usepackage{showkeys}\fi

\def\showkeysrefformat#1{{\normalfont\tiny\ttfamily#1}}
\makeatletter
\def\SK@@ref#1>#2\SK@{%
{\@inlabelfalse\leavevmode\vbox to\z@{%
\vss\SK@refcolor\rlap{\vrule\raise .75em%
\hbox{\showkeysrefformat{#2}}}}}}
\makeatother

\usepackage[a4paper,text={160mm,247mm},centering]{geometry}
\global\arraycolsep=2pt

\allowdisplaybreaks[3]

\numberwithin{equation}{section}

\expandafter\def\expandafter\bfseries\expandafter{\bfseries\ifmmode\else\boldmath\fi}
\expandafter\def\expandafter\mdseries\expandafter{\mdseries\ifmmode\else\unboldmath\fi}
\expandafter\def\expandafter\normalfont\expandafter{\normalfont\ifmmode\else\unboldmath\fi}

\def\g{\mathfrak g}
\def\e{\textrm{e}}
\def\tr{\textrm{tr}}
\def\mL{\mathcal{L}}
\def\Ad{\textrm{Ad}}

\def\beq{\begin{equation}}
\def\eeq{\end{equation}}
\def\beqz{\begin{equation*}}
\def\eeqz{\end{equation*}}
\def\bea{\begin{eqnarray}}
\def\eea{\end{eqnarray}}

 \def\ha{\mbox{\small $\frac{1}{2}$}}

\def\id{\protect{{1 \kern-.28em {\rm l}}}}

\def\co{\sigma}

\def\EL{\eta_{\scriptscriptstyle L}}
\def\ER{\eta_{\scriptscriptstyle R}}
\def\ELR{\eta_{\scriptscriptstyle L,R}}

\def\AL{\mathcal{A}_{\scriptscriptstyle L}}
\def\AR{\mathcal{A}_{\scriptscriptstyle R}}
\def\ALR{{\mathcal{A}}_{\scriptscriptstyle L,R}}

\def\OLL{O_{\scriptscriptstyle LL}}
\def\ORR{O_{\scriptscriptstyle RR}}
\def\OLR{O_{\scriptscriptstyle LR}}
\def\ORL{O_{\scriptscriptstyle RL}}

\def\OomLL{{O_{{\scriptscriptstyle LL},\omega}}}
\def\OomRR{{O_{{\scriptscriptstyle RR},\omega}}}
\def\OomLR{{O_{{\scriptscriptstyle LR},\omega}}}
\def\OomRL{{O_{{\scriptscriptstyle RL},\omega}}}

\def\OomLLt{{O_{{\scriptscriptstyle LL},\omega}^t}}
\def\OomRRt{{O_{{\scriptscriptstyle RR},\omega}^t}}
\def\OomLRt{{O_{{\scriptscriptstyle LR},\omega}^t}}
\def\OomRLt{{O_{{\scriptscriptstyle RL},\omega}^t}}

\def\OcalL{{\cal O}_{\scriptscriptstyle L}}
\def\OcalR{{\cal O}_{\scriptscriptstyle R}}
\def\OcalLR{{\cal O}_{\scriptscriptstyle L,R}}
 
\def\bEL{\EL}
\def\bER{\ER}
\def\bELR{\ELR}

\def\bk{k}

\def\ssLL{{\scriptscriptstyle LL}}

\def\gL{g_{\scriptscriptstyle L}}
\def\gR{g_{\scriptscriptstyle R}}
\def\gLR{g_{\scriptscriptstyle L,R}}

\def\gtL{\tilde g_{\scriptscriptstyle L}}
\def\gtR{\tilde g_{\scriptscriptstyle R}}
\def\gtLR{\tilde g_{\scriptscriptstyle L,R}}

\def\jpL{j_+^{\scriptscriptstyle L}}
\def\jpR{j_+^{\scriptscriptstyle R}}
\def\JpL{J_+^{\scriptscriptstyle L}}
\def\JpR{J_+^{\scriptscriptstyle R}}
\def\JpmL{J_\pm^{\scriptscriptstyle L}}
\def\jpmL{j_\pm^{\scriptscriptstyle L}}

\def\jmL{j_-^{\scriptscriptstyle L}}
\def\jmR{j_-^{\scriptscriptstyle R}}
\def\JmL{J_-^{\scriptscriptstyle L}}
\def\JmR{J_-^{\scriptscriptstyle R}}
\def\JpmR{J_\pm^{\scriptscriptstyle R}}
\def\jpmR{j_\pm^{\scriptscriptstyle R}}

\def\Jpm{J_\pm}

\def\JtpmL{{\tilde J}_\pm^{\scriptscriptstyle L}}
\def\JtpmR{{\tilde J}_\pm^{\scriptscriptstyle R}}

\def\wzk{\mbox{\tiny WZ} , \scriptscriptstyle k}

\def\ll{l}
\def\lmL{l_-^{\scriptscriptstyle L}}
\def\lmR{l_-^{\scriptscriptstyle R}}
\def\lpL{l_+^{\scriptscriptstyle L}}
\def\lpR{l_+^{\scriptscriptstyle R}}
\def\lpmL{l_\pm^{\scriptscriptstyle L}}
\def\lpmR{l_\pm^{\scriptscriptstyle R}}

\def\LmL{L_-^{\scriptscriptstyle L}}
\def\LmR{L_-^{\scriptscriptstyle R}}
\def\LpL{L_+^{\scriptscriptstyle L}}
\def\LpR{L_+^{\scriptscriptstyle R}}

\def\xL{x_{\scriptscriptstyle L}}
\def\xR{x_{\scriptscriptstyle R}}
\def\xLR{x_{\scriptscriptstyle L,R}}

\def\yL{y_{\scriptscriptstyle L}}
\def\yR{y_{\scriptscriptstyle R}}

\def\xtL{\tilde x_{\scriptscriptstyle L}}

\def\xhL{\hat x_{\scriptscriptstyle L}}
\def\xhR{\hat x_{\scriptscriptstyle R}}
\def\xhLR{\hat x_{\scriptscriptstyle L,R}}

\def\jtpL{\tilde \jmath_+^{\scriptscriptstyle L}}
\def\jtpR{\tilde \jmath_+^{\scriptscriptstyle R}}

\def\jtmL{{\tilde \jmath}_-^{\scriptscriptstyle L}}
\def\jtmR{{\tilde \jmath}_-^{\scriptscriptstyle R}}

\def\ww{\omega}
\def\wt{\omega^t}
\def\op{\mathcal{O}}

\def\zceL{\mbox{MC}_{\scriptscriptstyle L}}

\def\QL{{{\cal Q}_{\scriptscriptstyle L}}}
\def\QR{{{\cal Q}_{\scriptscriptstyle R}}}
\def\PL{{{\cal P}_{\scriptscriptstyle L}}}
\def\PR{{{\cal P}_{\scriptscriptstyle R}}}

\def\PLt{{{\cal P}_{\scriptscriptstyle L}^t}}
\def\PRt{{{\cal P}_{\scriptscriptstyle R}^t}}

\ifpublic

\else

\newcounter{comcompt}
\newcounter{piq}
\newcounter{treqle}
\newcounter{boxq}
\fi

\usepackage{fancyhdr}
\ifpublic\else
\fi

\makeatletter
\let\@keywords\@empty
\let\@subject\@empty
\providecommand{\keywords}[1]{\gdef\@keywords{#1}}
\providecommand{\subject}[1]{\gdef\@subject{#1}}
\def\thetitle{\@title}
\def\theauthor{\@author}
\def\thesubject{\@subject}
\def\thedate{\@date}
\def\thekeywords{\@keywords}
\makeatother
\AtBeginDocument{
\hypersetup{pdftitle={\thetitle}}
\hypersetup{pdfauthor={\theauthor}}
\hypersetup{pdfsubject={\thesubject}}
\hypersetup{pdfkeywords={\thekeywords}}}

\title{Combining the bi-Yang-Baxter deformation, the Wess-Zumino term and TsT transformations in one integrable \texorpdfstring{$\sigma$}{sigma}-model}
\author{F. Delduc, B. Hoare, T. Kameyama, M. Magro}

\begin{document}

\begin{center}

\vspace*{2cm}

\begingroup\Large\bfseries\thetitle\par\endgroup

\vspace{1.5cm}

\begingroup
F. Delduc$^*$\footnote{E-mail:~francois.delduc@ens-lyon.fr},
B. Hoare$^\dagger$\footnote{E-mail:~bhoare@itp.phys.ethz.ch},
T. Kameyama$^*$\footnote{E-mail:~takashi.kameyama@ens-lyon.fr},
M. Magro$^*$\footnote{E-mail:~marc.magro@ens-lyon.fr}
\endgroup

\vspace{1cm}

\begingroup
*\it Univ Lyon, Ens de Lyon, Univ Claude Bernard, CNRS, Laboratoire de Physique,
\\
F-69342 Lyon, France
\\
$^{\dagger}$\it Institut f\"ur Theoretische Physik, Eidgen\"ossische Technische Hochschule Z\"urich,
\\
Wolfgang-Pauli-Strasse 27, 8093 Z\"urich, Switzerland
\endgroup

\end{center}

\vspace{2cm}

\begin{abstract}
A multi-parameter integrable deformation of the principal chiral model is presented.
The Yang-Baxter and bi-Yang-Baxter $\sigma$-models, the principal chiral model plus
a Wess-Zumino term and the TsT transformation of the principal chiral model
are all recovered when the appropriate deformation parameters vanish.
When the Lie group is $SU(2)$, we show that this four-parameter integrable deformation
of the $SU(2)$ principal chiral model corresponds to the Lukyanov model.
\end{abstract}

\vfill

\pagebreak

\tableofcontents

 \section{Introduction} \label{sec intro}

In \cite{Lukyanov:2012zt} Lukyanov constructed a novel four-parameter integrable deformation of the $SU(2)$ principal chiral model (PCM), which preserves a $U(1) \times U(1)$ subgroup of the original $SU(2) \times SU(2)$ global symmetry.
This four-parameter model generalises \cite{Lukyanov:2012zt,Hoare:2014pna} a number of
previously well-known theories:
\begin{itemize}
\item Fateev's two-parameter deformation of the $SU(2)$ PCM \cite{Fateev:1996ea}.
This identification of the Fateev model as a special case of the Lukyanov model resolved
the long-standing question of the integrability of the Fateev model.
\item The $SU(2)$ PCM plus the Wess-Zumino (WZ) term with arbitrary coefficient \cite{Witten:1983ar}.
For a special value of this arbitrary coefficient one finds the conformal $SU(2)$
Wess-Zumino-Witten (WZW) model.
\item The TsT transformation of the $SU(2)$ WZW model, which can also be realised as a gauged WZW model for $(SU(2) \times U(1))/U(1)$ \cite{Horne:1991gn,Giveon:1991jj}.
\end{itemize}
Lukyanov's model is defined by a metric and $B$-field. In the undeformed limit,
the $B$-field vanishes and
the metric is the one of the three-sphere. One may then ask if the full four-parameter deformation
can be written as an action for a group-valued field $g \in SU(2)$, and in turn generalised
to arbitrary Lie group $G$.

\medskip

Our aim in this paper is to answer these questions.
To do this we will draw on a number of recent developments, many of which can
trace their origins to Klim\v{c}\'{\i}k's Yang-Baxter $\sigma$-model \cite{Klimcik:2002zj,Klimcik:2008eq}, a
one-parameter integrable deformation of the PCM for a general group $G$, whose appellation
reflects its dependence on a solution of the modified classical Yang-Baxter equation
for $\mathfrak{g} = \text{Lie}(G)$.

The Yang-Baxter $\sigma$-model can be generalised to a two-parameter integrable deformation
of the PCM, the bi-Yang-Baxter $\sigma$-model \cite{Klimcik:2008eq,Klimcik:2014bta}, which also incorporates the one-parameter Yang-Baxter deformation of the
symmetric space $\sigma$-model \cite{Delduc:2013fga} for cosets of the type $(G \times G)/G_{\text{diag}}$.
Algebraically the two parameters manifest as $q$-deformations of the $G \times G$
symmetry, with an independent deformation parameter for each factor of the group $G$
\cite{Delduc:2015xdm} (see also \cite{Kawaguchi:2011pf,Delduc:2013fga,Delduc:2016ihq}).

In \cite{Hoare:2014pna} it was shown that the bi-Yang-Baxter $\sigma$-model for $G = SU(2)$ is equivalent to Fateev's two-parameter deformation.
This model does not have a non-trivial coupling to the $B$-field.
In contrast the Lukyanov model does have such a coupling.
As discussed above, for a certain choice of parameters this $B$-field corresponds to a WZ term.
In \cite{Delduc:2014uaa} it was understood how to introduce such an anti-symmetric term for the Yang-Baxter $\sigma$-model while preserving classical integrability.
This construction of the Yang-Baxter deformation of the PCM plus WZ term has been achieved for any Lie group $G$ and
generalises the $SU(2)$ case \cite{Kawaguchi:2011mz,Kawaguchi:2013gma}.

The Yang-Baxter deformations of \cite{Klimcik:2002zj,Klimcik:2008eq,Delduc:2013fga} depend on a solution of the modified classical Yang-Baxter equation.
However, they can also be defined in terms of a solution of the classical Yang-Baxter equation 
\cite{Kawaguchi:2014qwa}. One of the simplest such solutions is when the $R$-matrix is 
abelian (i.e. when the generators from which it is built commute). In this case the homogeneous Yang-Baxter $\sigma$-model is equivalent to a TsT transformation \cite{Matsumoto:2014nra,Matsumoto:2015uja,Matsumoto:2014gwa,Matsumoto:2015jja,vanTongeren:2015soa,vanTongeren:2015uha,Osten:2016dvf}.

\medskip

In this paper we present a multi-parameter deformation of the PCM for a general group $G$ that incorporates each of the models introduced above.
We furthermore construct a Lax pair that encodes its equations of motion, thereby demonstrating the classical integrability of the model.
The number of deformation parameters depends on the group $G$.
For $G = SU(2)$ there are four parameters and in this case we explicitly demonstrate equivalence with Lukyanov's model \cite{Lukyanov:2012zt}.
Therefore, in this sense, the model is the generalisation of Lukyanov's model to arbitrary group $G$.

The construction of the model is split into two stages. In section \ref{sec 3def} we consider a 
three-parameter integrable model: the bi-Yang-Baxter deformation of the PCM plus WZ term, generalising the construction of \cite{Delduc:2014uaa}.
Generically this breaks the symmetry of the model from $G \times G$ to $U(1)^{\text{rank} \, G}
\times U(1)^{\text{rank} \, G}$, i.e. the Cartan subgroup.
We arrive at the Lagrangian and Lax pair for the multi-parameter deformation of the PCM in 
section \ref{sec TsT} by implementing a general TsT transformation that mixes the Cartan 
generators of the two copies of $G$, which provides $( {\text{rank} \, G})^2$ 
additional parameters. 
For $G = SU(2)$ the Cartan subgroup is one-dimensional and therefore there is one additional parameter.
In section \ref{sec Lukyanov} we demonstrate the equivalence to Lukyanov's model.
Finally we conclude in section \ref{sec conclusion} with comments and open questions.

\section{Bi-Yang-Baxter \texorpdfstring{$\sigma$}{sigma}-model plus WZ term}\label{sec 3def}

In this section we construct a three-parameter integrable deformation of the PCM.
Two of these parameters
correspond to those of the bi-Yang-Baxter $\sigma$-model while the third is
the coupling to the WZ term. To obtain this integrable deformation
of the PCM we employ on the following strategy.
First of all, we shall view the PCM for a Lie group $G$
as the $(G \times G)/G_{\text{diag}}$ symmetric space $\sigma$-model, where $G_{\text{diag}}$ is the
diagonal subgroup of $G \times G$. In the framework of integrable deformations, 
this perspective has been previously used in
\cite{Hoare:2014oua,Delduc:2015xdm}. Secondly, the $G_{\text{diag}}$ gauge invariance
will be realised by introducing a gauge field.
In subsection \ref{subsec 3def Action} we start from an ansatz for the action with five
free parameters and derive the corresponding equations of motion. We then determine the
conditions for this action to define an integrable field theory in subsection \ref{sec 3def Lax Pair}.
We show that a Lax pair exists provided the five parameters are fixed in terms of desired three
deformation parameters.

\subsection{Action}\label{ssec:3defact}

\label{subsec 3def Action}
Let $G$ be a semi-simple real Lie group. We shall start from the action
\bea \nonumber
S[\gLR, A] & = &
- \int d^2\co \, \sum_{a,b = L,R} \tr \left[ (j_+^a - A_+) O_{ab} (j_-^b - A_-) \right]
\\
&&
+ \, S_{\wzk}[\gL]-S_{\wzk}[\gR]- \bk \int d^2\co\,\tr\left[A_-(\jpL -\jpR)-A_+(\jmL-\jmR)\right],
\label{action2}
\eea
where $\co^\pm$ are light-cone coordinates.
The fields $\gL$ and $\gR$ take values in the Lie group $G$ while
the gauge field $A_\pm$ takes values in Lie algebra $\g$.
The left-invariant one-forms $j^{\scriptscriptstyle L}$ and $j^{\scriptscriptstyle R}$
are defined as $j^a=g_a^{-1}d g_a$ ($a = L,R$).
The operators $O_{ab}$ are given by
\bea \nonumber
\OLL & = & \Ad_{\gL}^{-1} \left[ (1+\bEL^2)\,\frac{1+\AL R}{1-\bEL^2R^2} \right] \Ad_{\gL} ,\qquad
\\ \nonumber
\ORR & = & \Ad_{\gR}^{-1} \left[ (1+\bER^2)\,\frac{1+\AR R}{1-\bER^2R^2} \right] \Ad_{\gR} ,
\\
\OLR& = &\ORL= 0,
\eea
with $\Ad_g(x) = g x g^{-1}$ for $x \in \g$. The operator $R$ is a
non-split $R$-matrix on $\g$. It is skew-symmetric and
solves the modified classical Yang-Baxter equation on $\g$, which means that for $x$ and $y$ in $\g$ we have
\begin{subequations}
\begin{align}
& \tr(x \,Ry \bigr) = - \tr(Rx \, y), \label{RSkewSym} \\
\left[Rx,Ry\right] & = R\bigl([Rx,y]+[x,Ry]\bigr)+[x,y]. \label{mCYBE}
\end{align}
\end{subequations}
Furthermore, we take $R$ to be a standard $R$-matrix, which implies that
\beq R^3 = -R , \label{rcubed} \eeq
and that its non-trivial kernel is the Cartan subalgebra $\mathfrak{h}$ of $\g$, i.e. 
\beq R x = 0 ,  \qquad \forall x \in \mathfrak{h} . \label{rkernel} \eeq
The term $S_{\wzk}$ in \eqref{subsec 3def Action} denotes the standard Wess-Zumino term,
\beq \label{WZ Term}
S_{\wzk}[g]=- k
\int d^2\co d\xi \, \tr \big[g^{-1} \partial_{\xi} g [g^{-1} \partial_+ g , g^{-1} \partial_- g] \big].
\eeq
The presence of the WZ term indicates that the associated coupling should be quantised in the quantum theory.
However, let us note that, in a mild abuse of notation, what we call $k$ is not the standard integer-valued level.

The action \eqref{action2} is invariant under $G_{\text{diag}}$ gauge transformations,
\beq \label{gauge}
\gLR \to \gLR g_0, \qquad A_\pm \to g_0^{-1} \partial_\pm g_0 + g_0^{-1} A_\pm g_0,
\eeq
with $g_0(\co^\pm)$ taking values in $G$. This is so because  
$O_{ab}$ transforms as $O_{ab} \to  \Ad_{g_0}^{-1} O_{ab} \Ad_{g_0}$ while 
$j^{\scriptscriptstyle L}_\pm - j^{\scriptscriptstyle R}_\pm$
and $j_\pm^{\scriptscriptstyle L,R} - A_\pm$  have the homogeneous transformations
$x \to \Ad_{g_0}^{-1} x$.

For the moment the coefficients $\ALR$ are free.
The way they depend on $\ELR$ and $k$
shall be fixed by imposing the existence of a Lax pair.
The resulting dependence coincides with the analogous
expressions in \cite{Kawaguchi:2011mz,Kawaguchi:2013gma,Delduc:2014uaa}.

Before we proceed to construct the Lax pair let us briefly illustrate the
motivation for using a gauge field. To determine a Lax pair, we will have to
explicitly invert operators such as $O_{ab}$. Without introducing a gauge
field, the $G_{\text{diag}}$ gauge invariance would be ensured by making use of
the projector onto the orthogonal complement of the diagonal subalgebra of $\g
\oplus \g$ (see e.g. \cite{Hoare:2014oua} for the bi-Yang-Baxter case). Such
insertions of the projector operator make inverting the relevant operators in a
tractable way substantially more difficult. As we shall see in the next
subsection, the presence of the gauge field thus allows the inversion to be
done in a simple way.

To construct a Lax pair we follow the method of \cite{Delduc:2014uaa} and
start by determining the equations of motion.
The equations of motion for the gauge field read
\beq \label{fieldeq}
\JpmL + \JpmR =0,
\eeq
where
\begin{subequations} \label{3def def J}
\bea
\JmL&=&(\OLL+k)(\jmL-A_-) ,\qquad
\JpL=(\OLL^t-k)(\jpL-A_+) , \label{3def def JL} \\
\JmR&=&(\ORR-k)(\jmR-A_-) ,\qquad
\JpR=(\ORR^t+k)(\jpR-A_+) .
\eea
\end{subequations}
In these expressions, the operators $\OLL^t$ and $\ORR^t$ are obtained by taking
the transpose of $\OLL$ and $\ORR$ respectively. This corresponds to flipping the sign of $R$.

The equations of motion for $\gL$ and $\gR$ are respectively given by
\begin{subequations} \label{3def eom gL gR}
\bea
D_+\JmL +D_- \JpL -2k F_{-+} &=0,\label{3def eom L}\\
D_+ \JmR +D_- \JpR +2k F_{-+}&=0.\label{3def eom R}
\eea
\end{subequations}
Here we have introduced
covariant derivatives $D_\pm x = \partial_\pm x + [A_\pm,x]$ and
$F_{-+}$ is the field strength of the gauge field,
\beqz
F_{-+}=\partial_-A_+-\partial_+A_-+[A_-,A_+].
\eeqz

\subsection{Lax Pair}\label{sec 3def Lax Pair}

To proceed we treat the equations of motion for the gauge field \eqref{fieldeq}
separately to those for $g_a$. In particular, as is typical for constraint
equations, they will not be determined by the zero curvature condition for the
Lax pair.

The currents $\JpmR$ can be obtained from the equations of motion
for the gauge field \eqref{fieldeq}, and are, on-shell, just the negative of $\JpmL$. We shall therefore
focus on the currents $\JpmL$, their equation of motion \eqref{3def eom L} and the
Maurer-Cartan equation,
\beq \label{ZCE L}
\partial_- \jpL -\partial_+ \jmL +[\jmL,\jpL]=0.
\eeq
From now on, we explicitly use the relation $R^3=-R$ in order to write all
operators, such as $\OLL$, as a linear combination of
$\Pi=1+R^2$, $R$ and $R^2$. The operator $\Pi$ is the projector on
the Cartan subalgebra $\mathfrak{h}$. To do this one can use the relations
\begin{subequations}
\begin{gather}
\Pi=1+R^2, \qquad \Pi^2=\Pi, \qquad \Pi R= R \Pi=0,\\
(a \Pi+bR+cR^2)^{-1}=a^{-1}\Pi+\frac{1}{b^2+c^2}(-bR+cR^2). \label{inv op formula}
\end{gather}
\end{subequations}
Now expressing the currents $\jpmL$ in terms of $\JpmL$ and $A$ using \eqref{3def def JL} and
\eqref{inv op formula} leads to
\beq
\jpmL = \left(a_\pm\Pi_\ssLL + b_\pm R _\ssLL + d_\pm R^2_\ssLL \right) \JpmL
+A_\pm \label{jL},
\eeq
where we make use of a general notation for operators dressed by the adjoint action, e.g.
$\Pi_\ssLL= \Ad_{\gL}^{-1} \Pi \Ad_{\gL}$.
The coefficients $a_\pm$, $b_\pm$ and $d_\pm$ are given by
\beq
a_\pm=\frac{1}{1+\EL^2\mp k}, \qquad
b_\pm=\frac{\pm\AL}{\AL^2+(1\mp k)^2},\qquad
d_\pm=-\frac{1\mp k}{\AL^2+(1\mp k)^2}.
\eeq
These coefficients satisfy the relations
\beq \label{3def relat param 0}
-b_+d_--b_-d_+= \ha( b_++b_-), \qquad
b_+b_--d_+d_-= \ha (d_++d_-).
\eeq
Note that the analogous expressions for the right currents
are obtained from the left ones by the replacement rule
$(L, \EL, \AL ,k) \rightarrow (R,\ER,\AR,-k)$.

Let us denote the
left-hand side of the Maurer-Cartan equation \eqref{ZCE L} as $\zceL$. Starting from
\eqref{jL} we may rewrite $\zceL$ as
\bea
\zceL&=&\Big(\frac{a_+-a_-}{2}\Pi_\ssLL+\frac{b_+-b_-}{2}R_\ssLL+\frac{d_+-d_-}{2}
R^2_\ssLL\Big)(D_- \JpL+D_+ \JmL)+F_{-+}\nonumber\\
&&+\Big(\frac{a_++a_-}{2}\Pi_\ssLL+\frac{b_++b_-}{2}R_\ssLL+\frac{d_++d_-}{2}R^2_\ssLL\Big)
(D_- \JpL -D_+ \JmL)\nonumber\\
&&+\Big(-(b_+b_-+d_+d_-+d_+a_-+a_+d_-)\Pi_\ssLL \nonumber
\\ && \phantom{ + \Big( } -(b_+d_-+d_+b_-)R_\ssLL
+(b_+b_--d_+d_-)R^2_\ssLL\Big)[\JmL,\JpL]\nonumber\\
&&\vphantom{\Big(}+b_+(a_-+d_-)\Pi_\ssLL[\JmL , R_\ssLL \JpL]+b_-(a_++d_+)\Pi_\ssLL[R_\ssLL \JmL , \JpL].
\label{flat1}
\eea
If we choose $\AL$ as in \cite{Kawaguchi:2011mz,Kawaguchi:2013gma,Delduc:2014uaa},
\beq \label{3def def AL}
\AL^2=\EL^2\Big(1-\frac{k^2}{1+\EL^2}\Big),
\eeq
then the coefficients $a_\pm$, $b_\pm$ and $d_\pm$ satisfy the following
relations
\begin{subequations}
\begin{align}
b_+(a_-+ d_-) = b_-(a_++d_+) & ,\label{3def relat param 1} \\
-(b_+b_-+d_+d_-+d_+a_-+a_+d_-) =\ha &(a_++a_-). \label{3def relat param 2}
\end{align}
\end{subequations}
This choice has the following consequences. Let us start with \eqref{3def relat param 1}.
Since the standard $R$-matrix satisfies (see for instance \cite{Delduc:2014uaa})
\beq
\Pi\big([Rx,y]+[x,Ry]\big)=0 \qquad \forall x,y \in \g,
\eeq
the last line in \eqref{flat1} vanishes. The next step is to use \eqref{3def relat param 2}
and \eqref{3def relat param 0} to combine the third line of \eqref{flat1}
with the second one. Finally, we use the equation of motion \eqref{3def eom L}
in the first line of \eqref{flat1}. Following these steps we obtain
\bea \nonumber
\zceL&=&\Big(\left[1+k(a_+-a_-)\right]\Pi_\ssLL + k(b_+-b_-)R_\ssLL+
\left[-1+k(d_+-d_-)\right]R^2_\ssLL \Big) F_{-+} \\
&&+\Big(\frac{a_++a_-}{2}\Pi_\ssLL + \frac{b_++b_-}{2}R_\ssLL
+\frac{d_++d_-}{2}R^2_\ssLL \Big) (D_-\JpL -D_+ \JmL +[\JmL , \JpL]). \qquad
\label{3def zce int1}
\eea
The condition \eqref{3def def AL} implies that the operators
appearing in the first and second lines of \eqref{3def zce int1} are proportional,
with the relative coefficient being equal to $(1+k^2+\AL^2)$. Furthermore,
these operators are invertible. Therefore, on-shell, the equation $\zceL=0$ is equivalent to
\beq \label{3def zceL int 2}
D_-\JpL - D_+ \JmL +[\JmL , \JpL] + (1+k^2+\AL^2)F_{-+} =0.
\eeq
Proceeding in the same way for the right currents, choosing in particular
\beq \label{3def def AR}
\AR^2=\ER^2\Big(1-\frac{k^2}{1+\ER^2}\Big),
\eeq
one similarly arrives at
\beq \label{3def zceR int 2}
D_-\JpR - D_+ \JmR + [\JmR , \JpR] + (1+k^2+\AR^2)F_{-+} =0.
\eeq

We now take the sum of \eqref{3def zceL int 2} and \eqref{3def zceR int 2} and use
the equations of motion for the gauge field \eqref{fieldeq} to express the field strength
$F_{-+}$ in terms of $[ \JmL , \JpL]$. We then use this expression for $F_{-+}$ in
\eqref{3def eom L} and \eqref{3def zceL int 2} to obtain
\begin{subequations} \label{3def almost Lax Pair}
\begin{align}
F_{-+} -\frac{1}{2k} (\alpha_+ - \alpha_-) [ \JmL , \JpL] =0 &,\\
D_+ \JmL +D_- \JpL -(\alpha_+ - \alpha_-)[\JmL , \JpL]&=0, \vphantom{\frac12} \\
D_+ \JmL -D_- \JpL -(\alpha_+ + \alpha_-)[\JmL , \JpL]&=0 \vphantom{\frac12} ,
\end{align}
\end{subequations}
with
\beq \label{def alphapm}
\alpha_+=\frac{-\AL^2+\AR^2-4k}{2(2(1+k^2)+\AL^2+\AR^2)},\qquad
\alpha_-=\frac{-\AL^2+\AR^2+4k}{2(2(1+k^2)+\AL^2+\AR^2)}.
\eeq
To construct a Lax pair let us redefine the gauge field as
\beq
\widehat{A}_\pm=A_\pm+\alpha_\pm \JpmL.
\eeq
The equations \eqref{3def almost Lax Pair} are then equivalent to
\begin{subequations} \label{3def very close Lax Pair}
\begin{align}
\widehat{F}_{-+}=F_{-+}-\alpha_+\alpha_-[\JmL , \JpL]&=-G^2[\JmL , \JpL],\\
\widehat{D}_+ \JmL +\widehat{D}_- \JpL&=0,\\
\widehat{D}_+ \JmL -\widehat{D}_- \JpL&=0,
\end{align}
\end{subequations}
where $\widehat{D}$ are covariant derivatives with respect to $\widehat{A}$ and
\beq \label{def G}
G^2=\frac{(4+(\AL+\AR)^2)(4+(\AL-\AR)^2)}{4(2(1+k^2)+\AL^2+\AR^2)^2}.
\eeq
The equations \eqref{3def very close Lax Pair} are equivalent to the flatness of the Lax pair
\beq \label{def Lax Pair 3def}
\mathcal{L}_\pm(\lambda)=\widehat A_\pm + G\,\lambda^{\pm1} \JpmL,
\eeq
where $\lambda$ is a spectral parameter. We have therefore shown that the action
\eqref{action2} defines an integrable model with $\ALR$ given by \eqref{3def def AL} and \eqref{3def def AR}.

\section{TsT transformation} \label{sec TsT}

The three-parameter deformation of the PCM constructed in section \ref{sec 3def} breaks the global $G \times G$ symmetry of the action.
As a consequence of the property \eqref{rkernel} the symmetry that remains is the Cartan subgroup specified by the kernel of the operator $R$.
By implementing TsT transformations \cite{Klimcik:1993kd,Horne:1991gn,Giveon:1991jj} on the corresponding shift isometries we are able to introduce additional deformation parameters while preserving integrability \cite{Frolov:2005ty,Frolov:2005dj,Alday:2005ww}.
In this section we perform a general TsT transformation with each of the two shift isometries coming from a different copy of $G$.

\subsection{On the action}\label{ssec:action}

Our starting point is the action \eqref{action2}.
As shown in subsection \ref{sec 3def Lax Pair} the equations of motion for $\gL$ and $\gR$ and the Maurer-Cartan equations follow from a Lax pair if $\AL$ and $\AR$ are fixed in terms of
$\bEL$, $\bER$ and $\bk$ as
\bea \label{alrt}
\AL^2 & = & \bEL^2 \Big(1 - \frac{\bk^2}{1+\bEL^2} \Big), \qquad
\AR^2 = \bER^2 \Big(1 - \frac{\bk^2}{1+\bER^2} \Big).
\eea
For $\bEL = \bER = 0$ the symmetry of the action is $G \times G$, which is broken to the Cartan subgroup $U(1)^{\text{rank}\,G} \times U(1)^{\text{rank}\,G}$ for generic values of the deformation parameters.
To implement the TsT transformations in the Cartan directions we start by making the corresponding shift isometries manifest.
To this end we parameterise ($a=L,R$)
\bea \label{paramgtgx}
g_a = \exp(x_a) \tilde g_a , \qquad \tilde g_a \in G , \quad x_a \in \mathfrak{h} ,
\eea
such that
\bea
j_\pm^a = \tilde \jmath_\pm^a + \Ad_{\tilde g_a}^{-1} \partial_\pm x_a ,
\eea
where $\tilde \jmath^a$ is the left-invariant one-form associated with $\tilde g_a$, i.e. $\tilde \jmath^a = \tilde g_a^{-1} d \tilde g_a$.
It is important to note that the parameterisation \eqref{paramgtgx} introduces a new left-acting Cartan gauge symmetry
\bea\label{cartangauge}
x_a \to x_a + \xi_a , \qquad \tilde g_a \to \exp(-\xi_a) \tilde g_a .
\eea
As we will see this symmetry survives the TsT transformation (up to potential total derivatives). Therefore for now we leave it unfixed, using the $x_a$ coordinates to implement the deformation, and fix it only at the end.

Defining the combinations
\bea
\ll_\pm^a = \Ad_{\tilde g_a} (\tilde \jmath_\pm^a - A_\pm) ,
\eea
which are invariant under the original $G_{\text{diag}}$ gauge transformations \eqref{gauge},
and the rescaled projections of $\ll_\pm^a$ onto the Cartan subalgebra (recall that $\Pi = 1 + R^2$ is the projector onto $\mathfrak{h}$)
\beq \label{Ldef}
L^{\scriptscriptstyle L}_\pm=(1+\EL^2\pm k) \, \Pi \, \lpmL , \qquad
L^{\scriptscriptstyle R}_\pm=(1+\ER^2\mp k) \, \Pi \, \lpmR ,
\eeq
we use \eqref{rcubed} and \eqref{rkernel} to rewrite the action \eqref{action2} in the form
\bea \nonumber
S[\gtLR, \xLR, A] & = &
- \int d^2\co \, \tr \big[ \lpL \, \OcalL \, \lmL + \LmL \partial_+ \xL +
\LpL \partial_- \xL + (1+\bEL^2) \partial_+ \xL \partial_- \xL \big]
\\ \nonumber
&& - \int d^2\co \, \tr \big[\lpR \, \OcalR \, \lmR +
\LmR \partial_+ \xR + \LpR \partial_- \xR + (1+\bER^2) \partial_+ \xR \partial_- \xR \big]
\\ \label{action3}
&& + \, S_{\wzk}[\gtL]-S_{\wzk}[\gtR]-\bk\int d^2\co\,\tr
\big[A_-(\jtpL- \jtpR)-A_+(\jtmL-\jtmR)\big], \qquad
\eea
where the operators $\OcalLR$ are given by
\bea
\OcalLR = 1 + \ALR R + \bELR^2 \Pi .
\eea
To implement the TsT transformation we first T-dualise $\xL \to \xtL$,
then perform the shift $\xR = \xhR + \ww \xtL$, where $\ww$ is a constant
linear operator on the Cartan subalgebra $\mathfrak{h}$ containing $(\text{rank} \,G)^2$ additional parameters,
and finally implement the reverse T-duality $\xtL \to \xhL$.
Eventually we arrive at the action
\bea \nonumber
S_\omega[\gtLR, \xhLR, A] & = &
- \int d^2\co \, \tr \big[\lpL \,\OcalL \,\lmL - (1+\bER^2) \LpL \wt \widetilde \op^{-1} \ww \LmL
\\ \nonumber && \phantom{+ \int d^2\co \, \tr \big[}
+ \LpL \op^{-1} \partial_- \xhL +
\LmL \op^{-1} \partial_+ \xhL + (1+\bEL^2) \partial_+ \xhL \op^{-1} \partial_- \xhL \big]
\\ \nonumber
&& - \int d^2\co \, \tr \big[\lpR \,\OcalR \, \lmR - (1+\bEL^2) \LpR \ww \op^{-1} \wt\LmR
\\ \nonumber && \phantom{+ \int d^2\co \, \tr \big[}
+\LpR \widetilde \op^{-1} \partial_- \xhR +
\LmR \widetilde \op^{-1} \partial_+ \xhR + (1+\bER^2) \partial_+ \xhR
\widetilde \op^{-1} \partial_- \xhR \big]
\\ \nonumber
&& + \int d^2\co \, \tr \big[(\LpL + (1+\bEL^2)\partial_+ \xhL) \op^{-1}
\wt (\LmR + (1+\bER^2)\partial_- \xhR)
\\ \nonumber
&& \phantom{+ \int d^2\co \, \tr \big[} -
(\LpR + (1+\bER^2)\partial_+ \xhR)
\widetilde \op^{-1} \ww
(\LmL + (1+\bEL^2)\partial_- \xhL)
\big]
\\
&&
+ \, S_{\wzk}[\gtL]-S_{\wzk}[\gtR]- \bk \int d^2\co\,\tr\big[A_-
(\jtpL - \jtpR)-A_+(\jtmL - \jtmR)\big], \qquad
\label{eq:actionwithx}
\eea
with
\bea \label{optildeop}
\op = 1+(1+\bEL^2)(1+\bER^2) \wt\ww , \qquad
\widetilde \op & = & 1+(1+\bEL^2)(1+\bER^2) \ww\wt .
\eea
Note that $\op$ and $\widetilde \op$ are related as follows
\beqz
\omega^t \widetilde \op^{-1} \omega = ( \omega^t \widetilde \op^{-1} \omega )^t
= \op^{-1} \omega^t \omega ,\qquad
\omega \op^{-1} \omega^t= ( \omega \op^{-1} \omega^t )^t
=\widetilde \op^{-1} \omega \omega^t .
\eeqz
In order to recast the action \eqref{eq:actionwithx} in a form generalising \eqref{action2} we parameterise
\bea \label{sgauge}
\tilde g_a = \exp(y_a) \hat g_a , \qquad \hat g_a \in G , \quad y_a \in \mathfrak{h}.
\eea
Setting
\bea \label{mgauge} y_{\scriptscriptstyle L} = - (1-\bk^2 \wt\ww)^{-1} (\xhL + \bk \ww^t \xhR) , \qquad
y_{\scriptscriptstyle R} = - (1-\bk^2 \ww \wt)^{-1} (\xhR + \bk \ww \xhL) , \eea
we find that the $\hat{x}_a$ dependence drops out of the action up to the total derivative
\bea \label{egauge}
-\bk^2\left[ \partial_+ \xhL (1-\bk^2 \wt\ww)^{-1} \ww^t \partial_- \xhR -
\partial_- \xhL (1-\bk^2 \wt\ww)^{-1} \wt \partial_+ \xhR \right] ,
\eea
which we also drop. We expect to be able to remove the dependence on $\hat x_a$ in this way as a consequence of the left-acting Cartan gauge invariance \eqref{cartangauge}. As foreseen this symmetry survives the TsT transformation up to potential total derivatives that we ignored in the T-dualisations.

Renaming $\hat g_a$ as $g_a$, we are finally left with the action
\bea \nonumber
S_{\omega}[\gLR, A] & = &
- \int d^2\co \, \sum_{a,b = L,R} \tr \left[ (j_+^a - A_+) O_{ab,\omega} (j_-^b - A_-) \right]
\\ \label{Action 3def omega bis}&&
+ \, S_{\wzk}[\gL]-S_{\wzk}[\gR]- \bk \int d^2\co\,\tr\left[A_-(\jpL- \jpR)-A_+(\jmL-\jmR)\right], \qquad
\eea
where the dressed operators are now given by
\bea \nonumber
\OomLL & = & \Ad_{\gL}^{-1} \big[ 1 + \AL R + \bigl(\bEL^2 - (1+\bER^2)(1+\bEL^2 + \bk)(1+\bEL^2 - \bk)
(\wt \widetilde \op^{-1}\ww ) \bigr) \Pi \big] \Ad_{\gL} ,
\\ \nonumber
\OomRR & = & \Ad_{\gR}^{-1} \big[ 1 + \AR R + \bigl(\bER^2 - (1+\bEL^2)(1+\bER^2 + \bk)(1+\bER^2 - \bk)
(\ww \op^{-1} \wt) \bigr) \Pi \big] \Ad_{\gR} , \nonumber
\\ \nonumber
\OomLR & = & \Ad_{\gL}^{-1} \big[ (1+\bEL^2+\bk) (1+\bER^2+\bk) (\op^{-1} \wt) \Pi \big] \Ad_{\gR} ,
\\ \label{def O omega}
\OomRL & = & - \Ad_{\gR}^{-1} \big[ (1+\bEL^2-\bk)(1+\bER^2-\bk) (\widetilde \op^{-1} \ww )
\Pi \big] \Ad_{\gL} ,
\eea
with $\ALR$ defined in terms of $\bELR$ and $\bk$ in \eqref{alrt}, $\op$ and $\widetilde \op$ given in \eqref{optildeop} and we recall that $\omega$ is an arbitrary constant linear operator on $\mathfrak{h}$.
As we will shortly demonstrate via the existence of a Lax pair this multi-parameter deformation of the PCM is integrable.

\medskip

Before we do so, let us briefly consider various limits of \eqref{Action 3def omega bis} in order to gain a better understanding of the model.
First we note that, as expected, upon setting $\omega = 0$ we recover the three-parameter deformation of section \ref{sec 3def}, i.e. the bi-Yang-Baxter deformation of the PCM plus WZ term.
Additionally setting either $\EL$ or $\ER$ to zero we expect to find the one-parameter Yang-Baxter deformation of the PCM plus WZ term constructed in \cite{Delduc:2014uaa}. The model of \cite{Delduc:2014uaa} depends on a single field $g \in G$ and hence to explicitly check this relation we integrate out the gauge field.
This is done in section \ref{ssec:elim} for the multi-parameter deformation \eqref{Action 3def omega bis}, with the resulting action given in \eqref{So}. The latter only depends on $\gL$ and $\gR$ through the combination $g = \gL \gR^{-1} $ as a consequence of the gauge symmetry \eqref{gauge}, and indeed setting $\omega = \ER = 0$ we recover the model of \cite{Delduc:2014uaa}.

It is also interesting to consider the limit $k = 0$, that is when the WZ term is no longer present.
In this case we can rewrite the deformed action in a form familiar in the context of Yang-Baxter deformations
\bea
S_\omega[\gLR, A]\Big|_{k = 0} = - \int d^2\co \, \tr \Big[ \begin{pmatrix} \jpL - A_+, & \jpR - A_+ \end{pmatrix} \cdot \mathscr{O} \cdot \begin{pmatrix} \jmL - A_-, & \jmR - A_- \end{pmatrix}^t \Big] ,
\eea
where the operator $\mathscr{O}$ is given by
\beq
\mathscr{O} =
\left( \begin{array}{cc}
\sqrt{1+ \bEL^2} &0\\
0& \sqrt{1+ \bER^2}
\end{array}
\right)
\cdot
\frac{1}{1-\mathscr{R}_{\gLR}}
\cdot
\left(
\begin{array}{cc}
\sqrt{1+ \bEL^2} &0\\
0& \sqrt{1+ \bER^2}
\end{array}
\right),
\eeq
which in turn is defined in terms of a linear operator $\mathscr{R}$ acting on $\g \oplus \g$
\bea \nonumber
\mathscr{R}_{\gLR} & = & \left(
\begin{array}{ccc}
\Ad_{\gL}^{-1} & & 0 \\
0 & & \Ad_{\gR}^{-1}
\end{array}
\right)
\cdot
\mathscr{R}
\cdot
\left(
\begin{array}{cc}
\Ad_{\gL} & 0 \\
0 & \Ad_{\gR}
\end{array}
\right),
\\
\mathscr{R} & = &
\left(
\begin{array}{cc}
\bEL R & \sqrt{(1+\bEL^2)(1+\bER^2)}\wt \Pi \\
- \sqrt{(1+\bEL^2)(1+\bER^2)} \ww \Pi & \bER R
\end{array}
\right).
\eea
For all $X = (\xL, \xR)^t$ and $Y = (\yL,\yR)^t$ in $\g \oplus \g$ the operator $\mathscr{R}$ satisfies the modified classical Yang-Baxter equation
\beq\label{mcybeext}
[ \mathscr{R} X, \mathscr{R} Y] - \mathscr{R}[ \mathscr{R}X, Y] - \mathscr{R}[X,\mathscr{R}Y] =
\left(
\begin{array}{c}
\bEL^2 [\xL,\yL]\\
\bER^2 [\xR,\yR]
\end{array}
\right).
\eeq
Note that the right-hand side of \eqref{mcybeext} is independent of $\omega$ and hence if we additionally set $\EL = \ER = 0$ the operator $\mathscr{R}$ satisfies the classical Yang-Baxter equation. In this case we are left with the homogeneous Yang-Baxter deformation of the PCM with an abelian $R$-matrix, which is equivalent to a series of TsT transformations \cite{Matsumoto:2014nra,Matsumoto:2015uja,Matsumoto:2014gwa,Matsumoto:2015jja,vanTongeren:2015soa,vanTongeren:2015uha,Osten:2016dvf}.
Alternatively we may set $\omega = 0$, in which case we recover the bi-Yang-Baxter sigma model of \cite{Klimcik:2008eq,Klimcik:2014bta,Hoare:2014oua,Delduc:2015xdm}.
Finally, if $\EL^2 = \ER^2 = \eta^2$ and 
$(\eta + \eta^{-1})^2 \omega\omega^t =  (\eta + \eta^{-1})^2 \omega^t\omega =1$ then the operator $\mathscr{R}$ satisfies
\beq
\eta^{-2} \mathscr{R}^2 = - 1.
\eeq
Therefore, $\eta^{-1} \mathscr{R}$ defines a complex structure on $G \times G$.
Yang-Baxter deformations based on complex structures have been explored in \cite{Bykov:2016pfu} and
typically give rise to particularly simple models.

\subsection{On the Lax Pair}\label{ssec:lax}

The Lax pair \eqref{def Lax Pair 3def}
for the three-parameter model described by the action \eqref{action2} is given by
\beq
\mL_\pm(\lambda)=A_\pm+\alpha_\pm \JpmL + G\lambda^{\pm 1} \JpmL ,\label{lax0}
\eeq
with the parameters $\alpha_\pm$, $G$ given in \eqref{def alphapm} and
\eqref{def G} respectively. The zero-curvature equation for $\mL_\pm(\lambda)$ implies
the equations of motion \eqref{3def eom gL gR} and Maurer-Cartan equations, \eqref{3def zceL int 2}
and \eqref{3def zceR int 2}.
Furthermore, it should be supplemented
with the equations of motion for the gauge field \eqref{fieldeq},
which are constraint equations fixing the gauge field in terms of the group fields.

The Lax pair \eqref{lax0} and the constraint equations \eqref{fieldeq} are
written in terms of the currents $J_\pm^a$ and the gauge field $A_\pm$, where
the dependence on $g_a$ is contained within the former. Therefore, to determine
the Lax pair and constraint equations for the TsT transformed model we
just implement the transformation on $\Jpm^a$, which gives the TsT transformed
currents. It then follows that the Lax pair for the model described by the TsT
transformed action \eqref{Action 3def omega bis} has the same form \eqref{lax0}
as it had before transformation, only now with $\Jpm^a$ given by TsT
transformed expressions for the currents. The same holds for the constraint
equations \eqref{fieldeq}.

\medskip

To construct the currents of the TsT transformed model we start from
those of the three-parameter model defined by \eqref{3def def J}.
Using the parameterisation \eqref{paramgtgx} these can be written as
\beq
\JpmL=\JtpmL+Ad_{\gtL}^{-1}(1\mp k+\EL^2)\partial_\pm \xL, \qquad
\JpmR=\JtpmR+Ad_{\gtR}^{-1}(1\pm k+\ER^2)\partial_\pm \xR, \label{modJ}
\eeq
where the $\tilde J^a_\pm$ are simply obtained from $J^a_\pm$
by the replacement $g_a\rightarrow\tilde g_a$.
The currents $J^a_\pm$, and thus the Lax
pair, equations of motion and Maurer-Cartan equations, only depend on
derivatives of the Cartan subalgebra
valued fields $x_a$. Then, following, for example, \cite{Frolov:2005dj}, we track
the fate of the derivatives $\partial_\pm x_a$
through the TsT transformation.

In the first step, that is under the T-duality $\xL \,\rightarrow \xtL$, one has
\beq \label{eqeq1}
\partial_\pm \xL=-\frac{1}{1+\EL^2}(L^{\scriptscriptstyle L}_\pm\mp\partial_\pm \xtL),
\eeq
where $L_\pm^a$ are defined in \eqref{Ldef}.
The second step is a translation of $\xR$ and implies
\beq \label{eqeq2}
\hat{x}_{\scriptscriptstyle R}=\xR -\omega \xtL \quad \Rightarrow \quad \partial_\pm
\xR=\partial_\pm \hat{x}_{\scriptscriptstyle R} +\omega\partial_\pm \xtL.
\eeq
Finally, the second T-duality, $\xtL \rightarrow \hat x_{\scriptscriptstyle L}$, gives
\beq\begin{array}{l} \label{eqeq3}
\partial_\pm \xtL =\op^{-1}\Bigl(\pm L^{\scriptscriptstyle L}_\pm-(1+\EL^2)(\omega^t
(L^{\scriptscriptstyle R}_\pm +
(1+\ER^2)\partial_\pm \hat x_{\scriptscriptstyle R})\pm\partial_\pm\hat x_{\scriptscriptstyle L})\Bigr),
\end{array}\eeq
where $\op$ is defined in \eqref{optildeop}.
Once the TsT transformation is performed,
we fix the gauge $\hat x_{\scriptscriptstyle L}=
\hat x_{\scriptscriptstyle R}=0$ using the gauge symmetry \eqref{cartangauge}.
Recall that, as discussed in subsection \ref{ssec:action}, this symmetry survives the
TsT transformation up to total derivatives, which do not contribute
to the equations of motion.
For this gauge choice the expressions for $\partial_\pm x_a$ in \eqref{eqeq1} and \eqref{eqeq2} become
\beq
\begin{array}{l}
\partial_\pm \xL=-(1+\ER^2)\op^{-1}\omega^t \omega
L^{\scriptscriptstyle L}_\pm\mp \op^{-1}\omega^t L^{\scriptscriptstyle R}_\pm,\cr
\partial_\pm \xR=-(1+\EL^2)\tilde \op^{-1}\omega\omega^t
L^{\scriptscriptstyle R}_\pm \pm \tilde \op^{-1}\omega L^{\scriptscriptstyle L}_\pm.
\end{array}
\eeq
Substituting into \eqref{modJ} we find expressions for TsT transformed currents $\Jpm^a$
as a function of the field $\tilde g_a \in G$. Finally, to match with
the action \eqref{Action 3def omega bis} we rename
$\tilde g_a$ as $g_a$, after which these currents are expressed as follows
\bea
\JmL &=&(\OomLL+k)(\jmL-A_-)+\OomLR(\jmR-A_-),\nonumber\\
\JpL&=&(\OomLLt - k)(\jpL - A_+)+\OomRLt(\jpR-A_+),\nonumber\\
\JmR&=&(\OomRR-k)(\jmR-A_-)+\OomRL(\jmL-A_-),\nonumber\\
\JpR&=&(\OomRRt+k)(\jpR-A_+)+\OomLRt(\jpL-A_+), \label{J in terms of j A}
\eea
where the various operators are defined in \eqref{def O omega}. Therefore, the
Lax pair of the TsT transformed model \eqref{Action 3def omega bis} takes the
form \eqref{lax0} with $\Jpm^a$ now given by \eqref{J in terms of j A}. As
before, this Lax pair should be supplemented by constraint equations of the
form \eqref{fieldeq}, again with $\Jpm^a$ given by \eqref{J in terms of j A}.
Note that these results also follow from direct computation, in the spirit of
subsection \ref{sec 3def Lax Pair}, starting from the action \eqref{Action 3def
omega bis}.

\subsection{Elimination of the gauge field}\label{ssec:elim}

Let us now eliminate the gauge field from
the action \eqref{Action 3def omega bis}. The resulting action
will be the starting point in the next section for the
comparison with Lukyanov's model.

The equations of motion for the gauge field \eqref{fieldeq} and the definitions of
$J_\pm^a$ given in \eqref{J in terms of j A}
can be used to write the left-invariant currents
$j_\pm^a$ as
\bea \nonumber
\jmL -A_-&=&\QL \JmL ,\qquad
\jmR -A_-=-\QR \JmL,\\
\jpL -A_+&=&\PLt \JpL ,\qquad
\jpR -A_+=-\PRt \JpL,
\eea
where $\QL$, $\QR$, $\PL$ and $\PR$ are the following
operators
\bea
\nonumber
\QL&=&\bigl(\OomLL +k- \OomLR ( \OomRR -k)^{-1} \OomRL \bigr)^{-1}
\bigl(1+\OomLR (\OomRR -k)^{-1}\bigr),\\
\nonumber
\QR&=&\left(\OomRR -k-\OomRL(\OomLL+k)^{-1}\OomLR\right)^{-1}\left(1+\OomRL
(\OomLL+k)^{-1}\right),\\
\nonumber
\PL&=&\left(1+(\OomRR+k)^{-1}\OomRL \right)\left(\OomLL-k-\OomLR(\OomRR+k)^{-1}
\OomRL\right)^{-1},\\
\PR&=&\left(1+(\OomLL-k)^{-1}\OomLR \right)\left(\OomRR+k-\OomRL (\OomLL - k)^{-1}\OomLR\right)^{-1}.
\label{qlrplr}
\eea
Inverting these relations it is then possible to express the gauge field as
\beq \label{resol A eom}
A_-=\ha \left[\jmL + \jmR -\left(\QL - \QR \right) \JmL \right],\qquad
A_+=\ha \left[\jpL + \jpR -\left(\PLt - \PRt\right)\JpL \right].
\eeq
and the currents $\JpmL$ as
\beq
\JmL=\left(\QL + \QR \right)^{-1}j_-,\qquad
\JpL=\left(\PLt+\PRt\right)^{-1}j_+,
\label{Jpm}
\eeq
where
\beq
j_\pm=\jpmL-\jpmR.
\eeq
These results enable us to rewrite the first term in the Lagrangian for the action
\eqref{Action 3def omega bis} as
\bea \nonumber
&& -\sum_{a,b = L,R} \tr \left[ (j_+^a - A_+) O_{ab,\omega} (j_-^b - A_-) \right]
\\
&& \qquad \qquad = - \ha \, \tr\left[j_+\left(\QL+ \QR \right)^{-1}j_- + j_+ \left(\PL + \PR \right)^{-1}j_-\right].
\eea
The last term of \eqref{Action 3def omega bis} is proportional to the gauge field.
We can therefore use the relation \eqref{resol A eom} to obtain
\bea\nonumber
-k\,\tr\left[A_-(\jpL - \jpR)-A_+(\jmL - \jmR)\right]
&=& -k\,\tr\left[\jpL\jmR -\jpR \jmL \right] \\
&& \nonumber +\ha k\,\tr\left[j_+\left(\QL - \QR \right)\left(\QL + \QR \right)^{-1}j_-\right] \\
&& -\ha k\,\tr\left[j_+\left(\PL + \PR \right)^{-1}\left(\PL - \PR \right)j_-\right]. \label{L2}
\eea
The first term in \eqref{L2} may be combined with the WZ terms
associated with $\gL$ and $\gR$ using the Polyakov-Wiegmann formula \cite{Polyakov:1983tt}
\beq
S_{\wzk}[\gL]-S_{\wzk}[\gR]-k\int d^2\co\,\tr\left[\jpL \jmR -\jpR \jmL \right]=S_{\wzk}[\gL \gR^{-1}].
\eeq

Summing all these contributions gives
\bea
S_\omega[g = \gL\gR^{-1}] & = &
S_{\wzk}[g] - \ha \int d^2\co\,\tr\Big[
j_+\big(1-k(\QL - \QR )\big) (\QL + \QR )^{-1}j_-\nonumber\\
&& \phantom{S_{\wzk}[g] - \ha \int d^2\co\,\tr\Big[} +j_+(\PL + \PR )^{-1} \big(1+k(\PL - \PR )\big) j_-
\Big], \qquad
\label{So}
\eea
where the operators $\QL$, $\QR$, $\PL$ and $\PR$ are defined in
\eqref{qlrplr}. It is straightforward to check that, as indicated, this action
only depends on $\gL$ and $\gR$ through the combination $g = \gL \gR^{-1}$.
This is expected as a consequence of the $G_{\text{diag}}$ gauge symmetry
\eqref{gauge}.

\section{Equivalence with the Lukyanov model for \texorpdfstring{$G = SU(2)$}{G = SU(2)}} \label{sec Lukyanov}

In this section we prove that the action \eqref{Action 3def omega bis}
corresponds to the Lukyanov model \cite{Lukyanov:2012zt} for $G = SU(2)$. Let
us start by noting that $SU(2)$ has rank one. Therefore in this case the operator $\omega$,
introduced in section \ref{sec TsT}, contains just a single
parameter. In a slight abuse of notation we will also call this parameter $\omega$,
with the operator given by multiplying by the identity (acting on the Cartan subalgebra).
For $G=SU(2)$, the action \eqref{Action 3def omega bis}
thus defines a four-parameter integrable deformation of the $SU(2)$
PCM. As a first order check of equivalence we observe that this is the
same number of deformation parameters as in the Lukyanov
model.

To demonstrate the full equivalence we shall start with the action \eqref{So},
obtained after eliminating the gauge field.
Partial identification of this four-parameter deformation with the Lukyanov
model, that is to say with some deformation parameters set to zero, has already
been shown in \cite{Hoare:2014pna}. For this reason we use the same
parameterisation of $g \in SU(2)$ as in \cite{Hoare:2014pna}. We then compute
the corresponding metric and $B$-field and show that there exists a coordinate
transformation, and a map between the parameters $\AL$, $\AR$, $k$, $\omega$ and
Lukyanov's parameters $\kappa$, $p$, $h$ and $\bar h$, such that this metric and
$B$-field coincides with those of
\cite{Lukyanov:2012zt}.

\medskip

Let us take the $SU(2)$ group element
\beq
g(r,\phi,\psi)=\e^{-T^3 (\phi + \psi)}\left(r \, \id-2 \sqrt{1 - r^2}\,T^1\right)\e^{-T^3 (\phi - \psi)}.
\eeq
Here $T^i$ are the generators of $\mathfrak{su}(2)$ satisfying
\beq
[T^i,T^j]=\epsilon^{ij}{}_{k}T^k,\qquad
\tr(T^iT^j)=-\frac{1}{2}\,\delta^{ij}, \qquad i,j,k = 1,2,3,
\eeq
where the totally anti-symmetric tensor $\epsilon^{ijk}$ is normalised as $\epsilon^{123} = +1$
and the $\mathfrak{su}(2)$ indices are raised and lowered by $\delta^{ij}$ and its inverse.
The $R$-matrix acts on the generators as
\beq
R(T^+)=-iT^+,\qquad R(T^-)=iT^-,\qquad R(T^3)=0,
\eeq
where $T^\pm=\frac{1}{\sqrt{2}}(T^1\pm iT^2)$.

The computation of the metric and $B$-field is rather lengthy but ultimately straightforward.
In order to see the equivalence with the metric and $B$-field of Lukyanov's model
one needs to perform the following coordinate transformations for the angle
variables $\phi$ and $\psi$
\bea
\phi&=&\chi_1+\frac{f_{{\scriptscriptstyle L}}^++f_{\scriptscriptstyle R}^+ }{4(1-k\omega) (1 + \bEL^2)
(1 + \bER^2) \AL \AR}\log\left[\frac{4 + (\AL - \AR)^2}{4 + (\AL - \AR)^2+ 4 r^2 \AL \AR}\right],\nonumber\\
\psi&=&-\chi_2-\frac{f_{{\scriptscriptstyle L}}^- -f_{\scriptscriptstyle R}^-}{4 (1+k\omega)(1 + \bEL^2)
(1 + \bER^2) \AL \AR}\log\left[\frac{4 + (\AL - \AR)^2}{4 + (\AL - \AR)^2+ 4 r^2 \AL \AR}\right],
\label{angle-coord-transf}
\eea
where $f_{\scriptscriptstyle L,R}^\pm$ are given by
\bea
f_{{\scriptscriptstyle L}}^\pm&=&\AL (1 + \bEL^2)(k \bER^2\pm(1+\bER^2)(1+\bER^2-k^2)\omega) ,\nonumber\\
f_{\scriptscriptstyle R}^\pm&=&\AR (1 + \bER^2)(k\bEL^2\pm(1+\bEL^2)(1+\bEL^2-k^2)\omega).
\eea
The resulting metric becomes block diagonal, i.e. $g_{r\chi_1}=g_{r\chi_2}=0$.
We also introduce a new radial coordinate $z$ related to $r$ by
\beq
r=\sqrt{\frac{(1-\kappa)(1+z)}{2(1-\kappa\, z)}},
\eeq
and define Lukyanov's parameters $(\kappa,p,h,\bar h)$ as
\bea
\kappa&=&\frac{\sqrt{4+(\mathcal{A}_L+\mathcal{A}_R)^2}-\sqrt{4+(\mathcal{A}_L-\mathcal{A}_R)^2}}{\sqrt{4+(\mathcal{A}_L+\mathcal{A}_R)^2}+\sqrt{4+(\mathcal{A}_L-\mathcal{A}_R)^2}},\qquad
p^2=-\frac{\bEL^2(1+\bER^2)\AR}{\bER^2(1+\bEL^2)\AL},
\nonumber\\
h_\pm&=&h\pm\bar h=-\frac{4H_\pm}{(\sqrt{4+(\mathcal{A}_L+\mathcal{A}_R)^2}
+\sqrt{4+(\mathcal{A}_L-\mathcal{A}_R)^2})H_0},
\label{idtfy}
\eea
where the quantities $H_0$, $H_+$ and $H_-$ are given by
\bea
H_0&=&\sqrt{(1+\bEL^2)(1+\bER^2)}(1-k^2\omega^2),\nonumber\\
H_+&=&
k\sqrt{\bEL^2\bER^2+\omega^2(1+\bEL^2)(1+\bER^2)\left(4+\AL^2+\AR^2+\omega^2(1+\bEL^2-k^2)(1+\bER^2-k^2)\right)},
\nonumber\\
H_-&=&\omega(k^2+(1+\bEL^2)(1+\bER^2)).
\eea
With these identifications we indeed recover the metric and $B$-field of \cite{Lukyanov:2012zt}.
In particular, up to a total derivative, the Lagrangian corresponding to \eqref{So} is
\cite{Lukyanov:2012zt, Hoare:2014pna}
\bea
L &=&T\Bigl[U(z)\,\partial_+z\,\partial_-z+D(z)\,\partial_+\chi_1\,\partial_-\chi_1+\hat{D}(z)\,\partial_+\chi_2\,\partial_-\chi_2\nonumber\\
&&+\left[C(z)+B(z)\right]\partial_+\chi_1\,\partial_-\chi_2+\left[C(z)-B(z)\right]\partial_+\chi_2\,\partial_-\chi_1
\Bigr],
\label{Lukyanov}
\eea
where we have rewritten the Lukyanov background using new angle variables $(\chi_1,\chi_2)$,
related to the original ones $(v,w)$ through
$\chi_1=\ha R^{-1}(v-w)$, $\chi_2=\ha (v+w)$ \cite{Hoare:2014pna}.
The overall factor $T$ is equal to
\beq
T=\frac{2((1+\bEL^2)(1+\bER^2)+k^2)}{2+\bEL^2+\bER^2},
\eeq
while the components of \eqref{Lukyanov} are
\begin{eqnarray}
U(z)&=&\frac{m^2}{4(1-z^2)(1-\kappa^2z^2)},\nonumber\\
D(z)&=&R^2(1+z)\left[2+\kappa(p^2+p^{-2})-\kappa(2\kappa+p^2+p^{-2})z\right]Q(z),\nonumber \vphantom{\frac12} \\
\hat{D}(z)&=&(1-z)\left[2+\kappa(p^2+p^{-2})+\kappa(2\kappa+p^2+p^{-2})z\right]Q(z), \vphantom{\frac12}
\nonumber\\
C(z)&=&\kappa(p^2-p^{-2})R(1-z^2)Q(z),\nonumber \vphantom{\frac12} \\
B(z)&=&- \frac{2R\,m}{c+\bar{c}}\left[h(c^2-1)(\bar{c}-z)-\bar{h}(\bar{c}^2-1)(c+z)\right]Q(z)
,\label{componentL}
\end{eqnarray}
where $Q(z)$ is given by
\beq
Q(z)=\frac{(c+1)(\bar{c}-1)}{4(1-\kappa^2)(c+z)(\bar{c}-z)}.
\eeq
Finally, we recall the definitions of $c$, $\bar c$, $m$ and $R$,
\bea \nonumber
c&= &\sqrt{\frac{1+h^2}{\kappa^2+h^2}},\qquad
\bar c = \sqrt{\frac{1+\bar h^2}{\kappa^2+\bar h^2}},
\\
m& = &\sqrt{(\kappa+p^2)(\kappa+p^{-2})},
\qquad
R = \sqrt{\frac{(c-1)(\bar{c}+1)}{(c+1)(\bar{c}-1)}}. \label{c cbar m R}
\eea
The expressions of $c$, $\bar c$ and $m$ in terms of the parameters
$\AL$, $\AR$, $k$ and $\omega$ are cumbersome. We shall therefore not reproduce them here.
However, let us point out that the relations \eqref{idtfy} and \eqref{c cbar m R}
lead to a simple expression for $R$,
\beq
R=\frac{1-k\omega}{1+k\omega}.
\label{R}
\eeq
This expression is interesting because it has simple limits. Indeed, we have $R=1$
when $\omega=0$ or $k=0$.
This result is consistent with those obtained previously in \cite{Hoare:2014pna}.

\section{Conclusion} \label{sec conclusion}

In this paper we have presented a new multi-parameter integrable deformation of the PCM for a general group $G$.
The first step of its construction was the derivation of the integrable bi-Yang-Baxter deformation of the PCM plus WZ term.
The second was the implementation of a general TsT transformation mixing the Cartan generators of the two copies of $G$.

This multi-parameter integrable model generalises Lukyanov's four-parameter deformation of the $SU(2)$ PCM \cite{Lukyanov:2012zt} to arbitrary group $G$.
Therefore the construction confirms the proposal of \cite{Hoare:2014pna} on the algebraic origin of the four parameters: two correspond to the bi-Yang-Baxter deformation, one parameterises the coupling to the WZ term, and the final parameter is generated by a TsT transformation.

There are a number of possible open questions whose investigation would further probe the properties of this integrable $\sigma$-model.
One of the most important is the study of its classical Poisson structure, Hamiltonian integrability and twist function in the spirit of \cite{Delduc:2013fga,Delduc:2015xdm,Delduc:2016ihq,Vicedo:2015pna}.
To determine the twist function, it is enough to consider the three-parameter case.
Indeed, the twist function is not changed under a TsT transformation \cite{Vicedo:2015pna}.
One particular aim is to understand the $q$-deformed algebra of hidden charges.
Furthermore, the extension to the affine algebra as considered in \cite{Delduc:2017brb}
for the Yang-Baxter $\sigma$-model would be interesting to investigate
(see also \cite{Kawaguchi:2012ve,Kawaguchi:2012gp,Kawaguchi:2013gma} for the $SU(2)$ case). 
Finally, studying this $\sigma$-model  at the Hamiltonian level would indicate if it is also 
possible to reinterpret it as a dihedral affine Gaudin model \cite{Vicedo:2017cge}.

In \cite{Klimcik:2017ken} the Yang-Baxter deformation of the PCM plus WZ term of \cite{Delduc:2014uaa} was recast in the framework of $\mathcal{E}$-models \cite{Klimcik:2015gba,Klimcik:2016rov} (a first-order action defined on the Drinfel'd double). Understanding how to formulate the bi-Yang-Baxter deformation of the PCM plus WZ term (and TsT transformations thereof) presented in sections \ref{sec 3def} (and \ref{sec TsT}) in this language may prove useful in gaining a deeper understanding of the underlying algebraic structure of the model.

Setting $\EL = \ER = \omega = 0$ and $k = 1$, the deformed action simplifies to the WZW action for the group $G$. This model is conformal, as are its deformations associated with TsT transformations.
It would be interesting to investigate which other points in parameter space correspond to conformal sigma models at the quantum level and hence define string backgrounds.
In particular, this would involve generalising the one-loop renormalisation analysis, including UV and IR fixed points, of \cite{Lukyanov:2012zt} beyond the $SU(2)$ case.

Finally, there are a class of superstring backgrounds for which the Green-Schwarz worldsheet action takes the form (at least in part) of an integrable supercoset $\sigma$-model \cite{Metsaev:1998it,Berkovits:1999zq,Bena:2003wd,Zarembo:2010sg,Wulff:2015mwa}.
For the maximally symmetric $AdS_5 \times S^5$ background the $PSU(2,2|4)/(SO(1,4)\times SO(5))$ supercoset model of \cite{Metsaev:1998it} captures the full Green-Schwarz string.
Generalising the bosonic construction of \cite{Delduc:2013fga}, the Yang-Baxter deformation of this model was constructed in \cite{Delduc:2013qra,Delduc:2014kha}.

Particularly relevant to the constructions of this paper are string backgrounds for which the superisometry takes the form of a product group.
For example, the $AdS_3 \times S^3 \times T^4$ background is related to the supercoset $PSU(1,1|2)^2/(SU(1,1) \times SU(2))_{\text{diag}}$, and the $AdS_3 \times S^3 \times S^3 \times S^1$ background to the supercoset $D(2,1;\alpha)^2/(SU(1,1) \times SU(2) \times SU(2))_{\text{diag}}$ \cite{Babichenko:2009dk}.
In these cases one can construct a bi-Yang-Baxter deformation of the supercoset $\sigma$-model \cite{Hoare:2014oua}, or alternatively introduce a WZ term \cite{Cagnazzo:2012se}.
The results presented in this paper provide the first steps towards combining these two constructions into a single three-parameter model, on top of which further parameters may be introduced via TsT transformations.

\paragraph{Acknowledgments.}

We thank S. Lacroix and B. Vicedo for comments on the draft. The work of 
FD, TK and MM is partially supported by the French Agence Nationale de la Recherche (ANR) under grant ANR-15-CE31-0006 DefIS.
The work of BH is partially supported by grant no. 615203 from the European Research 
Council under the FP7.

\providecommand{\href}[2]{#2}\begingroup\raggedright\endgroup

\end{document}